\documentclass[]{JHEP3}
\usepackage{amssymb}
\usepackage{amsmath}

\title{Background-independent charges in Topologically Massive Gravity}

\author{Olivera Mi\v{s}kovi\'{c} $^{a,b}$ and Rodrigo Olea $^{a}$\\

$^a$Instituto de F\'{\i}sica, Pontificia Universidad
Cat\'{o}lica de Valpara\'{\i}so,\\
Casilla 4059, Valpara\'{\i}so, Chile.\\
$^b$Max-Planck-Institut f\"{u}r Gravitationsphysik, Albert-Einstein-Institut,\\
Am M\"{u}hlenberg 1, 14476 Golm, Germany.\\
E-mail: \email{olivera.miskovic@ucv.cl}, \email{rodrigo.olea@ucv.cl}}

\preprint{AEI-2009-091}

\abstract{We construct background-independent Noether charges in Topologically Massive
Gravity with negative cosmological constant using its first-order formulation.

The procedure is carried out by keeping track of the surface terms in the
variation of the action, regardless the value of the
gravitational Chern-Simons coupling $\mu $. In particular, this method provides a
definition of conserved quantities for solutions at the chiral point $\mu
\ell =1$ ($\ell $ is the AdS radius) that contain logarithmic terms (Log
Gravity).

It is also shown that the charge formula gives a finite result for
warped AdS black holes without the need for any background-substraction
procedure.}


\begin{document}



\section{Introduction}

Standard Einstein-Hilbert gravity in three spacetime dimensions is
topological in the sense that it does not possess local degrees of freedom.
Inclusion of a gravitational Chern-Simons term in the action gives rise to
propagating gravitons in an inequivalent theory known as Topologically
Massive Gravity (\textbf{TMG}) \cite{DJT}.

Nevertheless, it is commonly expected that the addition of higher-derivative
terms to the standard gravity action would render the theory unstable.
Indeed, in an anti-de Sitter context, the theory exhibits gravitons of
negative mass around AdS$_{3}$ background. In case of flat space, one has
the possibility of changing the sign of the Newton constant $G$, what turns
the theory unitary. However, for gravity with negative cosmological
constant, this trick would also make negative the energy of BTZ black holes
and, thus, it cannot be performed \cite{CDWW,Deser}.

The renewed interest in TMG with cosmological constant has been mainly
sparked by the claim that the theory becomes stable and chiral for a special
value of the topological mass parameter $\mu $, such that $\mu \ell =1$,
where $\ell $ is the AdS radius \cite{LSS}. It has been argued that, at this
\emph{chiral point}, the left-moving central charge $c_{L}$ of the boundary
conformal algebra vanishes and the local degree of freedom that corresponds
to a negative mass graviton becomes a pure-gauge state. This idea presents a
new perspective on the problem of achieving a unitary theory in AdS$_{3}$,
especially in the light of the recent Witten's proposal \cite{Witten,MW}.

However, several authors provided evidence against this conjecture,
manifested as the existence of left-moving excitations at the chiral point
\cite%
{CDWW,Grumiller-Johansson1,Park,CDWW2,GJJ,Carlip08,GKP,Blagojevic-CvetkovicTMG}%
, but that obey a relaxed version of Brown-Henneaux boundary conditions \cite%
{Grumiller-Johansson1,Grumiller-Johansson2,HMT}. In addition, an exact
solution of TMG compatible with Grumiller-Johansson fall-off conditions, but
not Brown-Henneaux ones was explicitly shown in ref.\cite%
{Garbarz-Giribet-Vasquez}.

Therefore, given the above results, the only possibility of chiral gravity
conjecture to be true is by considering from the beginning Brown-Henneaux
boundary conditions as part of the setup \cite{Maloney-Song-Strominger}.

Nevertheless, the extensive discussion on the topic was extremely
enlightening to learn about the subtleties of the boundary conditions in the
context of AdS$_{3}$/CFT$_{2}$ correspondence.

In the gauge/gravity duality framework, in order to extract the holographic
data of the theory, usually a regularized Brown-York stress tensor is
required \cite{Balasubramanian-Kraus}. The same procedure is used to define
background-independent charges in the gravity theory. However, it is
well-known that, due to the presence of higher-derivative terms in TMG there
is no analog of Gibbons-Hawking term to define the Dirichlet variational
problem for the boundary metric $h_{ij}$. As a consequence, a quasilocal
stress-tensor cannot be read off directly from the variation of the action
as $T^{ij}=\left( 2/\sqrt{-h}\right) (\delta I/\delta h_{ij})$. Only for
asymptotically AdS spacetimes, a holographic stress tensor can be obtained
considering a Fefferman-Graham-type expansion of the metric \cite%
{Kraus-Larsen,Solodukhin} (see \cite{Skenderis-Taylor-van Rees} for the
inclusion of leading and subleading logarithms in the metric).

On the other hand, the construction of charges is important to achieve the
proof of positivity of energy and stability of the solutions, that remains
as an open question. Background-substraction approaches to conserved
quantities \cite{Blagojevic-CvetkovicTMG,ADT,Clement} may be useful to deal
with this issue, but are not particularly insightful about the problem of
holography. Taking this into consideration, we derive here
background-independent charge formulas for TMG in view of a possible
holographic description in the non asymptotically AdS case.


\section{Obtaining TMG from first-order formalism\label{TMG-first}}

In first-order formalism, Topologically Massive Gravity with negative
cosmological constant $\Lambda =-1/\ell ^{2}$ is described by the action
\begin{equation}
I=-\frac{1}{16\pi G}\int\limits_{M}\epsilon _{ABC}\left( R^{AB}+\frac{1}{%
3\ell ^{2}}\,e^{A}e^{B}\right) e^{C}+\frac{1}{32\pi G\mu }%
\,\int\limits_{M}\left( L_{\text{CS}}(\omega )+2\lambda _{A}T^{A}\right)
+\int\limits_{\partial M}B\,,  \label{TMG}
\end{equation}%
where $M$ is a three-dimensional manifold with local coordinates $x^{\mu}$, $%
G$ is the (positive) gravitational constant, $\mu $ is a constant parameter
with dimension of mass and $L_{\text{CS}}$ is the gravitational Chern-Simons
3-form%
\begin{equation}
L_{\text{CS}}(\omega )=\omega ^{AB}d\omega _{BA}+\frac{2}{3}\,\omega _{\
B}^{A}\omega _{\ C}^{B}\omega _{\ A}^{C}\,.  \label{LCSw}
\end{equation}%
The fundamental fields of the theory are the dreibein $e^{A}=e_{\mu
}^{A}\,dx^{\mu }$ and the spin connection $\omega ^{AB}=\omega _{\mu
}^{AB}\,dx^{\mu }$, which define the curvature 2-form $R^{AB}=\frac{1}{2}%
\,R_{\mu \nu }^{AB}\,dx^{\mu }dx^{\nu }=d\omega ^{AB}+\omega _{\ \
C}^{A}\omega ^{CB}$ and the torsion 2-form $T^{A}=\frac{1}{2}\,T_{\mu \nu
}^{A}\,dx^{\mu }dx^{\nu }=De^{A}$. The covariant derivative acts on a
Lorentz vector $V^{A}$ as $DV^{A}=dV^{A}+\omega _{\ \ B}^{A}V^{B}$. For
simplicity, we have omitted wedge products ${\tiny {\wedge }}$ between the
differential forms.

The boundary term $B$ will be discussed in the next section.

The introduction of a Lagrange multiplier 1-form $\lambda _{A}(x)$
implements a torsionless condition on the manifold and consistently recovers
the dynamics of TMG \cite%
{Carlip08,Blagojevic-CvetkovicTMG,Deser-Xiang,Carlip91}.

Indeed, an arbitrary variation of the fields in the above action produces
the equations of motion
\begin{eqnarray}
\delta I &=&-\frac{1}{16\pi G}\int\limits_{M}\delta \omega ^{AB}\left[
\epsilon _{ABC}T^{C}+\frac{1}{\mu }\,\left( R_{AB}+\lambda _{A}e_{B}\right) %
\right]  \notag \\
&&-\frac{1}{16\pi G}\int\limits_{M}\delta e^{A}\left[ \epsilon _{ABC}\left(
R^{BC}+\frac{1}{\ell ^{2}}\,e^{B}e^{C}\right) -\frac{1}{\mu }\,D\lambda _{A}%
\right]  \notag \\
&&-\frac{1}{16\pi G\mu }\int\limits_{M}\delta \lambda
_{A}T^{A}+\int\limits_{\partial M}\Theta (\delta e,\delta \omega )\,,
\label{eom}
\end{eqnarray}
where the surface term is
\begin{equation}
\Theta =-\frac{1}{16\pi G}\,\epsilon _{ABC}\,\delta \omega ^{AB}e^{C}+\frac{1%
}{32\pi G\mu }\,\left( \delta \omega ^{AB}\omega _{BA}+2\delta e^{A}\lambda
_{A}\right) +\delta B\,.  \label{Theta B}
\end{equation}

The field equations then read
\begin{eqnarray}
T^{A} &=&0\,,  \label{T=0} \\
\frac{1}{\mu }\,\left[ R^{AB}+\frac{1}{2}\left( \lambda ^{A}e^{B}-\lambda
^{B}e^{A}\right) \right] +\epsilon ^{ABC}T_{C} &=&0\,,  \label{omega} \\
\epsilon _{ABC}\left( R^{BC}+\frac{1}{\ell ^{2}}\,e^{B}e^{C}\right) -\frac{1%
}{\mu }\,D\lambda _{A} &=&0\,.  \label{e}
\end{eqnarray}%
From eqs. (\ref{T=0}) and (\ref{omega}), the Lagrange multiplier can be
solved in terms of the Schouten tensor of the manifold%
\begin{equation}
S_{\mu \nu }=\left( Ric\right) _{\mu \nu }-\frac{1}{4}\,\mathcal{G}_{\mu \nu
}\,R  \label{Schouten}
\end{equation}%
as%
\begin{equation}
\lambda _{\mu }^{A}=-2\,e^{A\nu }S_{\mu \nu }\,.  \label{lambda}
\end{equation}%
The manifold is endowed with a spacetime metric $\mathcal{G}_{\mu \nu }=\eta
_{AB}\,e_{\mu }^{A}e_{\nu }^{B}$.

The relation (\ref{e}) together with eq.(\ref{lambda}) leads to the usual
field equation of TMG in second-order formalism,%
\begin{equation}
R_{\mu \nu }-\frac{1}{2}\,\mathcal{G}_{\mu \nu }\,R-\frac{1}{\ell ^{2}}\,%
\mathcal{G}_{\mu \nu }+\frac{1}{\mu }\,C_{\mu \nu }=0\,.
\end{equation}%
Here $C_{\ \nu }^{\mu }=\frac{1}{\sqrt{-\mathcal{G}}}\,\epsilon ^{\mu \alpha
\beta }\nabla _{\alpha }S_{\beta \nu }$ denotes the Cotton tensor obtained
from varying the gravitational Chern-Simons term with respect to the metric.
In our conventions, the Levi-Civita tensor density $\epsilon ^{\mu \alpha
\beta }$ is defined as $\epsilon ^{tr\phi }=+1$.


\section{Noether charges}

The bulk theory described by eq.(\ref{TMG}) is invariant under
diffeomorphisms $\delta x^{\mu }=\xi ^{\mu }(x)$ in the sense that the
equations of motion remain covariant, even though the full action is not
invariant.

For an action $I=\int_{M}L$ in terms of a Lagrangian $L=\frac{1}{3!}\,L_{\mu
\nu \lambda }\,dx^{\mu }dx^{\nu }dx^{\lambda }$, the corresponding Noether
current $J^{\mu }$ has the form%
\begin{equation}
^{\ast }J=\Theta \left( \pounds _{\xi }\omega ^{AB},\pounds _{\xi
}e^{A}\right) -I_{\xi }L\,,  \label{*J}
\end{equation}%
where the star $^{\ast }$ denotes the Hodge dual and $I_{\xi }L=\frac{1}{2}%
\,\xi ^{\mu }L_{\mu \nu \lambda }\,dx^{\nu }dx^{\lambda }$ is the
contraction operator acting on $L$. The symbol $\pounds _{\xi }$ stands for
the Lie derivative, which acts on the spin connection and vielbein as%
\begin{eqnarray}
\pounds _{\xi }\omega _{\mu }^{AB} &=&\partial _{\mu }\xi ^{\nu }\omega
_{\nu }^{AB}+\xi ^{\nu }\partial _{\nu }\omega _{\mu }^{AB}  \notag \\
&=&D_{\mu }\left( \xi ^{\nu }\omega _{\nu }^{AB}\right) +\xi ^{\nu }R_{\nu
\mu }^{AB}\,,  \label{LieOmega}
\end{eqnarray}%
and%
\begin{eqnarray}
\pounds _{\xi }e_{\mu }^{A} &=&\partial _{\mu }\xi ^{\nu }e_{\nu }^{A}+\xi
^{\nu }\partial _{\nu }e_{\mu }^{A}  \notag \\
&=&D_{\mu }\left( \xi ^{\nu }e_{\nu }^{A}\right) +\xi ^{\nu }T_{\nu \mu
}^{A}-\left( \xi ^{\nu }\omega _{\nu }^{AB}\right) e_{B\mu }\,,
\label{LieVielbein}
\end{eqnarray}%
respectively.

Generally speaking, the conservation equation $d\left. ^{\ast }J\right. =0$
implies, by virtue of the Poincar\'{e} lemma, that the current can always be
written locally as the exterior derivative of certain quantity. It is the
procedure developed in this section which permits to write down the current
as $^{\ast }J=d\mathcal{Q}(\xi )$ globally and, therefore, to identify the
Noether charge for a Killing vector $\xi =\xi ^{\mu }\partial _{\mu }$.

For the purpose of construction of conserved quantities in TMG, we shall
consider a boundary term $B$ given by
\begin{equation}
B=\frac{1}{32\pi G}\,\epsilon _{ABC}\,\omega ^{AB}e^{C},  \label{BoundCS}
\end{equation}%
i.e., a half of the Gibbons-Hawking (\textbf{GH}) term. Supplementing the
Einstein-Hilbert action with $B$ sets a well-posed variational principle
when the extrinsic curvature is held fixed at the boundary, and makes the
action finite for asymptotically AdS spacetimes \cite{Miskovic-Olea3D}. This
point can be seen by performing an explicit comparison to
Balasubramanian-Kraus regularization \cite{Balasubramanian-Kraus} (see
Appendix \ref{AdS group}). The unusual factor in eq.(\ref{BoundCS}) respect
to the one carried by the GH term arises naturally in the gauge formulation
of standard gravity Lagrangian as a single Chern-Simons density for $SO(2,2)$
group, and it is the simplest case of a regularization scheme known as
Kounterterm method \cite{OleaKTs}.

It is expected that any modification of the \textit{effective }cosmological
constant due to the Chern-Simons coupling $\mu $ would leave the form of $B$
unchanged, what should not be the case of the counterterm coupling in ref.%
\cite{Balasubramanian-Kraus}.

Plugging in eq.(\ref{BoundCS}) into the general form of the surface term (%
\ref{Theta B}), one finds%
\begin{equation}
\Theta =-\frac{1}{32\pi G}\,\epsilon _{ABC}\,\left( \delta \omega
^{AB}e^{C}-\omega ^{AB}\delta e^{C}\right) +\frac{1}{32\pi G\mu }\,\left(
\delta \omega ^{AB}\omega _{BA}+2\delta e^{A}\lambda _{A}\right) \,.
\label{theta}
\end{equation}

Splitting the Lagrangian as%
\begin{equation}
L=L_{1}+\frac{1}{32\pi G\mu }\left( L_{\text{CS}}(\omega )+2\lambda
_{A}T^{A}\right) \,,  \label{L}
\end{equation}%
where $L_{1}$ is the Einstein-Hilbert part Lagrangian with the boundary term
in eq.(\ref{BoundCS}), makes clear that the total charge $Q(\xi )$ is the
sum of two contributions%
\begin{equation}
Q(\xi )=Q_{1}(\xi )+Q_{2}(\xi )\,.\label{Q tot}
\end{equation}%
The first term, $Q_{1}(\xi )$,\ is\ associated to the standard gravity part
\cite{Miskovic-Olea3D}
\begin{equation}
Q_{1}(\xi )=-\frac{1}{32\pi G}\int\limits_{\Sigma _{\infty }}\epsilon
_{ABC}\,\left( \xi ^{\mu }\omega _{\mu }^{AB}e_{\nu }^{C}+\xi ^{\mu }e_{\mu
}^{A}\omega _{\nu }^{BC}\right) dx^{\nu }\,,  \label{Q1_form}
\end{equation}%
where $\Sigma _{\infty }$ is the boundary of the spatial section at constant
time. The construction of the second term, proportional to $\mu ^{-1}$, $%
Q_{2}(\xi )$, is carried out below.

Since the torsion vanishes, the spin-connection $\omega _{\mu }^{AB}$\ is a
function of the dreibein $e_{\mu }^{A}$%
\begin{equation}
\omega _{\mu }^{AB}=-e^{B\alpha }\partial _{\mu }e_{\alpha }^{A}+\Gamma
_{\nu \mu }^{\lambda }\,e_{\lambda }^{A}e^{B\nu },  \label{torsionless-Omega}
\end{equation}%
where $\Gamma _{\nu \mu }^{\lambda }$ is the Christoffel symbol. In
Gauss-normal form of the metric,%
\begin{equation}
ds^{2}=\mathcal{G}_{\mu \nu }\,dx^{\mu }dx^{\nu
}=N^{2}(r)dr^{2}+h_{ij}(r,x)\,dx^{i}dx^{j}\,,  \label{Gaussian}
\end{equation}%
where $r$ is the Schwarzschild-like radial coordinate (the boundary is at $%
r\rightarrow \infty $), some of the components of the Christoffel symbol are
related to the extrinsic curvature $K_{ij}=\frac{1}{2N}\,h_{ij}^{\prime }$ as%
\begin{equation}
\Gamma _{ij}^{r}=-\frac{1}{N}\,K_{ij},\qquad \Gamma _{rj}^{i}=NK_{j}^{i}\,.
\label{Christ-Gauss}
\end{equation}%
Here, the prime denotes radial derivative $\partial /\partial r$.

The line element at the boundary of the spacetime in the frame (\ref%
{Gaussian}) can be, in turn, written in a time-like ADM form%
\begin{equation}
h_{ij}\,dx^{i}dx^{j}=-N_{\Sigma }^{2}(t)\,dt^{2}+\sigma _{\underline{m}%
\underline{n}}\left( N^{\underline{m}}dt+dy^{\underline{m}}\right) \left( N^{%
\underline{n}}dt+dy^{\underline{n}}\right) \,,  \label{ADMh}
\end{equation}%
generated by a unit normal $u_{i}=(-N_{\Sigma },\vec{0})$. It is clear that
for a constant-time slice, $\sigma _{\underline{m}\underline{n}}$ is the
metric on $\Sigma _{\infty }$.

For a given set of asymptotic Killing vectors $\{\xi ^{i}\}$, the charge (%
\ref{Q1_form}) expressed in terms of boundary tensors is%
\begin{equation}
Q_{1}(\xi )=\int\limits_{\Sigma _{\infty }}\sqrt{\sigma }dy\,\xi
^{j}q_{(1)j}^{i}\,u_{i}\,,  \label{Q3tens}
\end{equation}%
where the integrand reads%
\begin{equation}
q_{(1)j}^{i}=\frac{1}{16\pi G}\,\epsilon _{mk}\,\left( \delta
_{l}^{k}K_{j}^{m}+\delta _{j}^{k}K_{l}^{m}\right) \epsilon ^{il}\,,
\label{q1boundary}
\end{equation}%
and we have used the convention $\epsilon ^{rij}=-\epsilon ^{ij}$.

The Noether current for the second part in the Lagrangian (\ref{L}) uses the
corresponding surface term
\begin{equation}
\Theta _{(2)}=\frac{1}{32\pi G\mu }\,\left( \pounds _{\xi }\omega
^{AB}\omega _{BA}+2\pounds _{\xi }e^{A}\lambda _{A}\right) \,,
\end{equation}%
and the explicit expressions of Lie derivative on the fields, eqs.(\ref%
{LieOmega}) and (\ref{LieVielbein}), such that%
\begin{eqnarray}
^{\ast }J_{(2)} &=&\frac{1}{32\pi G\mu }\,\left( \partial _{\alpha }\xi
^{\mu }\omega _{\mu }^{AB}+\xi ^{\mu }\partial _{\mu }\omega _{\alpha
}^{AB}\right) \omega _{BA\beta }\,dx^{\alpha }\wedge dx^{\beta }  \notag \\
&&+\frac{1}{16\pi G\mu }\,\left( \partial _{\alpha }\xi ^{\mu }e_{\mu
}^{A}+\xi ^{\mu }\partial _{\mu }e_{\alpha }^{A}\right) \lambda _{A\beta
}\,dx^{\alpha }\wedge dx^{\beta }  \notag \\
&&-\frac{1}{32\pi G\mu }\,\left[ I_{\xi }L_{\text{CS}}(\omega )+\xi ^{\mu
}\left( \lambda _{A\mu }T_{\alpha \beta }^{A}-2\lambda _{A\alpha }T_{\mu
\beta }^{A}\right) dx^{\alpha }\wedge dx^{\beta }\right] \,,  \label{J2}
\end{eqnarray}%
where the contraction operator acting on the gravitational Chern-Simons term
produces%
\begin{equation}
I_{\xi }L_{\text{CS}}(\omega )=\xi ^{\mu }\left( \omega _{\mu }^{AB}\partial
_{\alpha }\omega _{BA\,\beta }-\omega _{\alpha }^{AB}\partial _{\mu }\omega
_{BA\,\beta }+\omega _{\alpha }^{AB}\partial _{\beta }\omega _{BA\,\mu
}+2\omega _{\mu \,B}^{A}\,\omega _{\alpha \,C}^{B}\,\omega _{\beta
\,A}^{C}\right) dx^{\alpha }\wedge dx^{\beta }.
\end{equation}

We show now that the different contributions in the current can be
rearranged as a total derivative plus the equations of motion in the form%
\begin{equation*}
^{\ast }J=d\mathcal{Q}+(\xi \cdot \omega ^{AB})(\text{e.o.m.})_{AB}+(\xi
\cdot e^{A})(\text{e.o.m.})_{A}+(\xi \cdot \lambda _{A})(\text{e.o.m.}%
)^{A}\,.
\end{equation*}

Indeed, the part proportional to $1/\mu $ is%
\begin{eqnarray}
J_{(2)}^{\lambda } &=&\frac{1}{64\pi G\mu }\frac{\,\epsilon ^{\lambda \alpha
\beta }}{\sqrt{-\mathcal{G}}}\left[ \partial _{\alpha }\left( \xi ^{\mu
}\omega _{\mu }^{AB}\omega _{BA\beta }+2\xi ^{\mu }e_{\mu }^{A}\lambda
_{A\beta }\right) \right.  \notag \\
&&\left. -\left( \xi ^{\mu }\omega _{\mu }^{AB}\right) \left( R_{BA\alpha
\beta }+2\lambda _{A\beta }e_{B\alpha }\right) -\left( \xi ^{\mu }\lambda
_{A\mu }\right) T_{\alpha \beta }^{A}-2\left( \xi ^{\mu }e_{\mu }^{A}\right)
D_{\alpha }\lambda _{A\beta }\right] \,.
\end{eqnarray}%
On the other hand, in the derivation of the first part of the Noether
charge, the corresponding current is
\begin{eqnarray}
J_{(1)}^{\lambda } &=&\frac{1}{64\pi G\mu }\,\frac{\epsilon ^{\lambda \alpha
\beta }}{\sqrt{-\mathcal{G}}}\left[ \partial _{\alpha }\left( -\epsilon
_{ABC}\,\left( \xi ^{\mu }\omega _{\mu }^{AB}e_{\beta }^{C}+\xi ^{\mu
}e_{\mu }^{A}\omega _{\beta }^{BC}\right) \right) \right.  \notag \\
&&\left. -\left( \xi ^{\mu }\omega _{\mu }^{AB}\right) \,\epsilon
_{ABC}T_{\alpha \beta }^{C}+2\left( \xi ^{\mu }e_{\mu }^{A}\right) \epsilon
_{ABC}\left( \frac{1}{2}\,R_{\alpha \beta }^{BC}+\frac{1}{\ell ^{2}}%
\,e_{\alpha }^{B}e_{\beta }^{C}\right) \right] ,
\end{eqnarray}%
what makes easy reading off the second piece of the Noether charge from the
first line of eq.(\ref{J2}),%
\begin{equation}
Q_{2}(\xi )=\frac{1}{32\pi G\mu }\int\limits_{\Sigma _{\infty }}\left( \xi
^{\mu }\omega _{\mu }^{AB}\omega _{BA\nu }+2\xi ^{\mu }e_{\mu }^{A}\lambda
_{A\nu }\right) dx^{\nu }.  \label{Q2 form_general}
\end{equation}

It can be shown that the Noether charge $Q(\xi)$ in eq.(\ref{Q tot}) is equivalent (up to finite
terms that do not modify the value of the mass and angular momentum) to a Hamiltonian
charge defined in ref.\cite{Iyer-Wald} associated to the same isometry $\xi$ (see Appendix \ref{Wald}).

In what follows, we first evaluate the charges derived in this section for
the Ba\~{n}ados-Teitelboim-Zanelli (\textbf{BTZ}) black hole \cite{BTZ,BHTZ}%
, obtained from a global AdS spacetime through identifications that preserve
the constant-curvature property. After that, we consider TMG solutions with
more general asymptotics.


\subsection{Solutions with constant curvature}

For spacetimes which are locally AdS, the Riemann tensor satisfies
\begin{equation}
R_{\mu \nu }^{\alpha \beta }=-\frac{1}{\ell ^{2}}\,\delta _{\left[ \mu \nu %
\right] }^{\left[ \alpha \beta \right] }\,,
\end{equation}%
what implies a particular value of the Lagrange multiplier (\ref{lambda}) in
terms of the dreibein as%
\begin{equation}
\lambda _{\mu }^{A}=\frac{1}{\ell ^{2}}\,e_{\mu }^{A}\,.
\label{lambda-constant}
\end{equation}%
Constant-curvature condition solves trivially the equations of motion of TMG.

In that case, the corresponding Noether charge is%
\begin{equation}
Q_{2}(\xi )=\frac{1}{32\pi G\mu }\int\limits_{\Sigma _{\infty }}\left( \xi
^{\mu }\omega _{\mu }^{AB}\omega _{BA\nu }+\frac{2}{\ell ^{2}}\,\xi ^{\mu
}e_{\mu }^{A}e_{A\nu }\right) dx^{\nu }.  \label{Q2ConstantCurv}
\end{equation}%
By virtue of the torsionless condition for the spin connection (\ref%
{torsionless-Omega}) and the Gaussian form of the Christoffel symbol (\ref%
{Christ-Gauss}), one can rewrite the charge (\ref{Q2ConstantCurv}) for
asymptotic Killing vectors in terms of quantities defined at the boundary,
in a similar form to eq.(\ref{Q3tens})%
\begin{equation}
Q_{(2)}(\xi )=\int\limits_{\partial \Sigma }\sqrt{\sigma }dy\,\xi
^{j}q_{(2)j}^{i}u_{i}\,,  \label{Q2}
\end{equation}%
where%
\begin{eqnarray}
q_{(2)j}^{i} &=&\frac{1}{32\pi G\mu }\left( \Gamma _{\nu j}^{\lambda }\Gamma
_{\lambda k}^{\nu }+\frac{2}{\ell ^{2}}\,h_{jk}\right) \frac{\epsilon ^{ik}}{%
\sqrt{-h}}  \notag \\
&=&\frac{1}{32\pi G\mu }\left( -2K_{jl}K_{\,k}^{l}+\Gamma _{mj}^{l}(h)\Gamma
_{lk}^{m}(h)+\frac{2}{\ell ^{2}}\,h_{jk}\right) \frac{\epsilon ^{ik}}{\sqrt{%
-h}}\,.  \label{q2boundary}
\end{eqnarray}%
We have dropped a few terms containing derivatives of the boundary zweibein
of the type $\partial _{i}e_{j}^{A}$. These are scheme-dependent
contributions, because they arise from the difference between $L_{\text{CS}%
}(\omega )$ in (\ref{LCSw}) and the gravitational Chern-Simons in terms of
the Christoffel symbol $L_{\text{CS}}(\Gamma )$%
\begin{equation}
L_{\text{CS}}(\Gamma )=\epsilon ^{\mu \nu \lambda }\left( \Gamma _{\mu
\alpha }^{\beta }\,\partial _{\nu }\Gamma _{\lambda \beta }^{\alpha }+\frac{2%
}{3}\,\Gamma _{\mu \alpha }^{\gamma }\,\Gamma _{\nu \beta }^{\alpha
}\,\Gamma _{\lambda \gamma }^{\beta }\right)
\end{equation}%
which can be expressed as a topological invariant plus a boundary term,%
\begin{equation}
L_{\text{CS}}(\Gamma )-L_{\text{CS}}(\omega )=\epsilon ^{\mu \nu \lambda }%
\left[ -\frac{1}{3}\left( e_{A}^{\alpha }\partial _{\mu }e_{\beta
}^{A}\right) \left( e_{B}^{\beta }\partial _{\nu }e_{\sigma }^{B}\right)
\left( e_{C}^{\sigma }\partial _{\lambda }e_{\alpha }^{C}\right) +\partial
_{\mu }(\omega _{\nu }^{AB}\partial _{\lambda }e_{B\alpha }e_{A}^{\alpha })%
\right] \,.
\end{equation}

The rotating solution of Einstein-Hilbert AdS gravity in three dimensions is
the BTZ black hole given by the metric%
\begin{equation}
ds^{2}=\mathcal{G}_{\mu \nu }\,dx^{\mu }dx^{\nu }=-f^{2}(r)dt^{2}+\frac{%
dr^{2}}{f^{2}(r)}+r^{2}\left( N_{\phi }(r)dt+d\phi \right) ^{2}\,,
\label{BTZ-like}
\end{equation}%
with the azimuthal angle $0\leq \phi \leq 2\pi $ and where the metric
function and angular shift read
\begin{equation}
f^{2}(r)=-8GM+\frac{r^{2}}{\ell ^{2}}+\frac{16G^{2}J^{2}}{r^{2}}\,,\qquad
N_{\phi }(r)=-\frac{4GJ}{r^{2}}\,.
\end{equation}%
The horizons of the BTZ black hole are the roots of the equation $f^{2}(r)=0$%
, that is,%
\begin{equation}
r_{\pm }^{2}=4GM\ell ^{2}\left( 1\pm \sqrt{1-\frac{J^{2}}{\ell ^{2}M^{2}}}%
\right) \,.
\end{equation}

Using eqs. (\ref{q1boundary}) and (\ref{q2boundary}) for the metric (\ref%
{BTZ-like}) give a formula for the total mass%
\begin{eqnarray}
\mathcal{M} &\equiv &Q(\partial _{t})=\lim\limits_{r\rightarrow \infty }%
\frac{1}{8G}\left[ \left( -f^{2}+\frac{r^{2}}{\ell ^{2}}+r^{2}N_{\phi
}^{2}\right) +\frac{2}{\mu }\,\frac{r^{2}N_{\phi }}{\ell ^{2}}\right] ,
\notag \\
&=&M-\frac{J}{\mu \ell ^{2}}\,\ ,
\end{eqnarray}%
whereas for the total angular momentum we obtain%
\begin{eqnarray}
\mathcal{J} &\equiv &Q(\partial _{\phi })=\lim\limits_{r\rightarrow \infty }%
\frac{1}{8G}\left[ -2r^{2}N_{\phi }+\frac{1}{\mu }\,\left( f^{2}-\frac{r^{2}%
}{\ell }-\,r^{2}N_{\phi }^{2}\right) \right] ,  \notag \\
&=&J-\frac{M}{\mu }.
\end{eqnarray}

The above Noether charges agree with the standard results computed by
canonical methods \cite{Garcia et al, ACL, Deser-Kanik-Tekin,
Olmez-Sarioglu-Tekin,Blagojevic-Cvetkovic} and holographic procedures \cite%
{Garbarz-Giribet-Vasquez,Kraus-Larsen,Solodukhin}, and they satisfy the
first law of black hole thermodynamics%
\begin{equation}
T\,\delta S=\delta \mathcal{M}+\Omega \,\delta \mathcal{J\,},
\label{first-law}
\end{equation}%
where $T$ is the Hawking temperature%
\begin{equation}
T=\frac{1}{4\pi }f^{\prime }(r_{+})=\frac{1}{2\pi r_{+}}\left( \frac{%
r_{+}^{2}}{\ell ^{2}}-\frac{16G^{2}J^{2}}{r_{+}^{2}}\right) \,,
\end{equation}%
and $\Omega $ is the angular velocity of the horizon%
\begin{equation}
\Omega =N_{\phi }(r_{+})=-\frac{r_{-}}{\ell r_{+}}\,.
\end{equation}%
The relation (\ref{first-law}) is valid only if one considers a contribution
to the entropy proportional to the inner horizon due to the gravitational
Chern-Simons term, that is,%
\begin{equation}
S=\frac{2\pi r_{+}}{4G}\left( 1-\frac{1}{\ell \mu }\frac{r_{-}}{r_{+}}%
\right) .
\end{equation}%
The general contribution to the entropy from the gravitational Chern-Simons
term was first computed by Solodukhin \cite{Solodukhin} using the conical
singularity method, which does not rely on the equations of motion. For BTZ
black hole, the correction to the macroscopic entropy due to the inner
horizon was derived from the first law of black hole thermodynamics in ref.%
\cite{Park-GCS}, where the Noether charges in first-order formalism were
also found. Solodukhin's general formula has been reobtained by means of
Tachikawa's procedure \cite{Tachikawa}, which incorporates corrections to
the Wald's formula due to the fact that $L_{CS}(\Gamma )$ is not invariant
under diffeomorphisms.

The form of the charges for the constant-curvature case is identical to the
one derived from a Lagrangian that is the linear combination of two
Chern-Simons densities for the $SO(2,2)$ group, constructed up with
invariant tensors of different parity (see Appendix \ref{AdS group})%
\begin{equation}
\tilde{L}=-\frac{\ell}{16\pi G}\left\langle AdA+\frac{2}{3}\,A^{3}\right\rangle
_{1}+\frac{1}{16\pi G\mu }\left\langle AdA+\frac{2}{3}\,A^{3}\right\rangle
_{2}\,.  \label{L CS}
\end{equation}%
Indeed, following the Noether procedure (see ref.\cite{Miskovic-Olea3D}),
one can obtain the Noether charge associated to a Killing vector $\xi ^{\mu
} $ for an arbitrary Chern-Simons action. For the above case, the conserved
quantity takes the form%
\begin{equation}
\tilde{Q}(\xi )=-\frac{\ell }{16\pi G}\int\limits_{\Sigma _{\infty }}\xi
^{\mu }\left\langle A_{\mu }A_{\nu }\right\rangle _{1}\,dx^{\nu }+\frac{1}{%
16\pi G\mu }\int\limits_{\Sigma _{\infty }}\xi ^{\mu }\left\langle A_{\mu
}A_{\nu }\right\rangle _{2}\,dx^{\nu }.
\end{equation}%
Substituting the traces for the generators of the AdS group defined in
Appendix \ref{AdS group}, eqs.(\ref{IT-epsilon}) and (\ref{IT-delta}), it is
straightforward to reproduce the charges (\ref{Q1_form}) and (\ref%
{Q2ConstantCurv}) from the corresponding pieces in the last expression.

The Lagrangian density (\ref{L CS}) induces a topological theory in four
dimensions which, in the conventions of the Appendix \ref{AdS group}, reads%
\begin{equation}
d\tilde{L}=-\frac{\ell }{16\pi G}\left\langle F^{2}\right\rangle _{1}+\frac{1%
}{16\pi G\mu }\left\langle F^{2}\right\rangle _{2}\,.  \label{dLtilde}
\end{equation}%
Defining the dual of the field strength in the indices of the universal
covering of AdS as%
\begin{equation}
\ast \mathcal{F}^{\underline{A}\underline{B}}=\frac{1}{2}\epsilon ^{%
\underline{A}\underline{B}\underline{C}\underline{D}}\mathcal{F}_{\underline{%
C}\underline{D}}\,,  \label{dualF}
\end{equation}%
the equation (\ref{dLtilde}) can be cast into the equivalent form%
\begin{eqnarray}
d\tilde{L} &=&-\frac{\ell }{64\pi G}\,\epsilon _{\underline{A}\underline{B}%
\underline{C}\underline{D}}\left( \mathcal{F}^{\underline{A}\underline{B}}%
\mathcal{F}^{\underline{C}\underline{D}}+\frac{1}{\mu \ell }\ast \mathcal{F}%
^{\underline{A}\underline{B}}\mathcal{F}^{\underline{C}\underline{D}}\right)
\,, \\
&=&-\frac{\ell }{32\pi G}\left\langle \left( F\pm \ast F\right)
^{2}\right\rangle _{1}\,-\frac{\ell }{16\pi G}\left( \frac{1}{\mu \ell }\mp
1\right) \,\left\langle F\ast F\right\rangle _{1}\,,  \label{self-dual mu}
\end{eqnarray}%
where, in the last line, one employs the identity%
\begin{equation}
\left\langle F^{2}\right\rangle _{1}=\frac{1}{2}\left( \left\langle
F^{2}\right\rangle _{1}+\left\langle \ast F^{2}\right\rangle _{1}\right) \,,
\end{equation}%
consistent with $\eta _{\underline{A}\underline{B}}=\,$diag$\left(
-1,1,1,-1\right) $. The form of eq.(\ref{self-dual mu}) makes clear that
when the Chern-Simons coupling is $\mu \ell =\mp 1$, $d\tilde{L}$ vanishes
identically for a globally (anti) self-dual AdS curvature $(\ast F=\pm F)$.
This fact resembles on the case of inclusion of topological invariants in
the four-dimensional AdS gravity action, where globally (anti) self-dual
solutions in the Weyl tensor are interpreted as the vacuum state of the
theory \cite{Miskovic-Olea4D}.


\subsection{Perturbed Extreme BTZ black hole\label{LogGravity}}

An axisymmetric solution of Topologically Massive Gravity at the chiral
point $\mu \ell =1$ has been found recently in ref.\cite%
{Garbarz-Giribet-Vasquez}. The Riemann tensor for this space is no longer
constant, because the metric contains a logarithmic branch. It is, however,
an asymptotically AdS spacetime defined in terms of the \emph{relaxed}
boundary conditions given by Grumiller and Johansson \cite%
{Grumiller-Johansson1,Grumiller-Johansson2}. Therefore, this is the first
exact stationary solution obtained for Log Gravity, as only pp-wave
solutions existed previously in the literature \cite{AyonBeato-Hassaine}.

For this solution, the line element in the ADM form is%
\begin{equation}
ds^{2}=-\mathcal{N}_{\perp }^{2}(r)\,dt^{2}+\frac{dr^{2}}{N^{2}(r)}%
+R^{2}(r)\left( d\phi -\mathcal{N}_{\phi }(r)dt\right) ^{2}\,,
\end{equation}%
with%
\begin{eqnarray}
\mathcal{N}_{\perp }^{2}(r) &=&N^{2}(r)-r^{2}N_{\phi
}^{2}(r)-N_{k}^{2}(r)+R^{2}(r)\mathcal{N}_{\phi }^{2}(r)\,, \\
R^{2}(r) &=&r^{2}+\ell ^{2}N_{k}^{2}(r)\,, \\
\mathcal{N}_{\phi }(r) &=&\frac{r^{2}N_{\phi }(r)+\ell N_{k}^{2}(r)}{R^{2}(r)%
}\,.
\end{eqnarray}%
The functions $N,N_{\phi }$ and $N_{k}$ in the above expressions are defined
as%
\begin{eqnarray}
N^{2}(r) &=&\frac{r^{2}}{\ell ^{2}}-8\pi GM+\frac{16\pi ^{2}G^{2}M^{2}\ell
^{2}}{r^{2}}\,, \\
N_{\phi }(r) &=&\frac{4\pi GM\ell }{r^{2}}\,, \\
N_{k}^{2}(r) &=&k\log \left( \frac{r^{2}-4\pi GM\ell ^{2}}{r_{0}^{2}}\right)
\,,  \label{log}
\end{eqnarray}%
where the constants $k$ and $r_{0}$ are two arbitrary parameters.

Summing the contributions of the charge formulas (\ref{Q1_form}) and (\ref%
{Q2 form_general}), the total mass and angular momentum are%
\begin{equation}
\mathcal{M}=\frac{k}{2G}\,,\qquad \mathcal{J}=-\frac{k\ell }{2G}\,.
\label{chargesGGV}
\end{equation}%
The dependence on the parameter $k$ makes evident the fact that, when the
logarithmic branch is switched off, the charges vanish identically, that
corresponds to the extreme BTZ black hole ($J=-M\ell $ in the conventions of
refs.\cite{LSS,Garbarz-Giribet-Vasquez}). The results in eq.(\ref{chargesGGV}%
) have been verified in different frameworks, which are the Clement's Super
Angular Momentum (\textbf{SAM}) method \cite{Clement-NMG} and the
Barnich-Brandt-Comp\`{e}re approach to conserved quantities \cite%
{Barnich-Brandt-Compere,Detournay-private}.

There is, however, a discrepancy in a global factor in the charges (\ref%
{chargesGGV}) respect to the ones computed in the original reference. As we
mentioned in the Introduction, TMG does not lend itself to a clear
definition of quasilocal stress tensor. This means that any attempt to
integrate by parts the variations of the extrinsic curvature $\delta K_{ij}=%
\frac{1}{2N}\partial _{r}\left( \delta h_{ij}\right) +\cdots$ and to cast
the variation of the action in the form $\delta I=\frac{1}{2}\int_{\partial
M}\sqrt{-h}\,\tau ^{ij}\delta h_{ij}$ would necessarily be ambiguous in the
definition of the new Brown-York\ stress tensor $\tau ^{ij}$. Thus, the
mismatch of (\ref{chargesGGV}) with the conserved quantities calculated in
ref.\cite{Garbarz-Giribet-Vasquez} must be due to this problem inherent to
TMG.


\subsection{Warped AdS$_{3}$ black holes}

It has been argued in ref.\cite{Anninos et al} that TMG could present stable
backgrounds which are spacelike, timelike or lightlike warped anti-de Sitter
spaces. In this work, the authors also conjecture the form of the
corresponding left and right moving central charges, simply based on an
entropy argument. The proposal for the value of $c_{L}$ was explicitly
verified from the asymptotic symmetry algebra in \cite{Compere-Detournay
central}.

It is well-known that BTZ black hole is obtained as the quotient of global
AdS$_{3}$ space by a discrete subgroup of $SL(2,\mathbb{R})\times SL(2,%
\mathbb{R})$ isometries. In a similar way, since warped AdS black holes are
locally warped AdS$_{3}$ with isometries $SL(2,\mathbb{R})\times U(1)$, they
are also constructed as identifications of warped AdS.

Spacelike (timelike) warped AdS black hole solutions of TMG are produced by
an identification along a Killing vector of a $SL(2,\mathbb{R})\times U(1)$-
invariant vacuum, where the $U(1)$ isometry is spacelike (timelike) \cite%
{Anninos et al}.

For convenience, the Chern-Simons coupling is redefined in terms of a
dimensionless parameter as $\nu =\mu \ell /3$.

For illustrative purposes, we compute here the conserved quantities
associated to the spacelike stretched black holes ($\nu ^{2}>1$). The ADM
form of the metric reads%
\begin{equation}
ds^{2}=-N^{2}(r)\,dt^{2}+\frac{\ell ^{2}dr^{2}}{4R^{2}(r)N^{2}(r)}%
+R^{2}(r)\left( d\phi +N_{\phi }(r)dt\right) ^{2}\,,  \label{ds2Warped}
\end{equation}%
where%
\begin{eqnarray}
R^{2}(r) &=&\frac{r}{4}\,\left[ 3r\left( \nu ^{2}-1\right) +\left( \nu
^{2}+3\right) \left( r_{+}+r_{-}\right) -4\nu \sqrt{r_{+}r_{-}\left( \nu
^{2}+3\right) }\right] \,,  \label{R2Warped} \\
N^{2}(r) &=&\frac{\left( \nu ^{2}+3\right) \left( r-r_{+}\right) \left(
r-r_{-}\right) }{4R^{2}(r)}\,, \\
N_{\phi }(r) &=&\frac{2\nu r-\sqrt{r_{+}r_{-}\left( \nu ^{2}+3\right) }}{%
2R^{2}(r)}\,,  \label{NphiWarped}
\end{eqnarray}%
and $\phi $ is the angle with a period $2\pi $.

Spacelike stretched black holes of this type were first obtained --in a
slightly different coordinate system-- in ref.\cite{Clement}.

In the general case, the integrand of the second part of the Noether charge (%
\ref{Q2 form_general}) can be expressed as%
\begin{equation}
q_{(2)j}^{i}=\frac{\ell }{96\pi G\nu }\left( -2K_{jl}K_{\,k}^{l}+\Gamma
_{mj}^{l}(h)\Gamma _{lk}^{m}(h)-4S_{jk}\right) \frac{\epsilon ^{ik}}{\sqrt{-h%
}}\,,
\end{equation}%
where $S_{jk}$ are the boundary components of the Schouten tensor (\ref%
{Schouten}),%
\begin{eqnarray}
S_{ij} &=&\mathcal{R}_{ij}(h)-\frac{1}{4}\,h_{ij}\,\mathcal{R}(h)-\frac{1}{N}%
\left( \partial _{r}K_{ij}-\frac{1}{2}\,h_{ij}\,\partial _{r}K\right)  \notag
\\
&&+2K_{ik}K_{j}^{k}-KK_{ij}+\frac{1}{4}\,h_{ij}\left(
K^{2}+K^{kl}K_{kl}\right) \,,
\end{eqnarray}%
as a consequence of Gauss-Codazzi relations listed in Appendix \ref%
{Gauss-Codazzi}. Here, $\mathcal{R}_{ij}(h)$ and $\mathcal{R}(h)$ are the
Ricci tensor and Ricci scalar of the boundary, respectively.

For a stationary metric described by eq.(\ref{ds2Warped}), the Schouten
tensor $S_{\mu }^{\nu }$ takes the general form found in Appendix \ref%
{Schouten Appix}. In particular, for warped AdS black holes with functions
eqs.(\ref{R2Warped}-\ref{NphiWarped}), it is given by%
\begin{equation}
S_{\nu }^{\mu }=\left(
\begin{array}{ccc}
-\frac{4\nu ^{2}-3}{2\ell ^{2}} & 0 & 0 \\
0 & \frac{2\nu ^{2}-3}{2\ell ^{2}} & 0 \\
-\frac{3\left( \nu ^{2}-1\right) \left( 2\nu r-\sqrt{r_{+}r_{-}\left( \nu
^{2}+3\right) }\right) }{2\ell ^{2}} & 0 & \frac{2\nu ^{2}-3}{2\ell ^{2}}%
\end{array}%
\right) .
\end{equation}%
In terms of $R(r)$, $N(r)$ and $N_{\phi }(r)$, the conserved quantities for
the metric (\ref{ds2Warped}) are%
\begin{eqnarray}
\mathcal{M} &=&-\frac{R}{8\ell G}\,\left( 2N^{2}R^{\prime }-R(N^{2})^{\prime
}+R^{3}(N_{\phi }^{2})^{\prime }\right) -  \notag \\
&&-\frac{R^{3}}{12\ell G\nu }\,\left[ N^{2}\left( 2\left( RN_{\phi }\right)
^{\prime \prime }+3R^{\prime }N_{\phi }^{\prime }\right) +3R^{3}(N_{\phi
}^{\prime })^{2}N_{\phi }\right.  \notag \\
&&\left. +(N^{2})^{\prime }(N_{\phi }R^{\prime }-\frac{1}{2}RN_{\phi
}^{\prime })-RN_{\phi }\left( N^{2}\right) ^{\prime \prime }\right]
\end{eqnarray}%
and%
\begin{equation}
\mathcal{J}=\frac{R^{3}}{4\ell G}\,\left[ \ell ^{2}RN_{\phi }^{\prime }+%
\frac{1}{3\nu }\left( 3\ell ^{2}R^{3}\left( N_{\phi }^{\prime }\right)
^{2}+2N\left( R^{\prime }N\right) ^{\prime }-R\left( N^{2}\right) ^{\prime
\prime }\right) \right] \,,  \label{J-Warped-functions}
\end{equation}%
that, substituting eqs.(\ref{R2Warped}-\ref{NphiWarped}), produces%
\begin{equation}
\mathcal{M}=\frac{(\nu ^{2}+3)}{48G\ell }\left( r_{+}+r_{-}-\frac{1}{\nu }%
\sqrt{r_{+}r_{-}\left( \nu ^{2}+3\right) }\right)  \label{Mass-Warped}
\end{equation}%
and
\begin{equation}
\mathcal{J}=\frac{\nu (\nu ^{2}+3)}{96G\ell }\left[ \left( r_{+}+r_{-}-\frac{%
1}{\nu }\sqrt{r_{+}r_{-}\left( \nu ^{2}+3\right) }\right) ^{2}-\frac{(5\nu
^{2}+3)}{4\nu ^{2}}(r_{+}-r_{-})^{2}\right] \,.  \label{J-Warped}
\end{equation}

Up to a factor $2$ in eq.(\ref{Mass-Warped}), the above quantities
correspond to the conserved quantities computed in ref.\cite{Anninos et al}
using the Abbott-Deser-Tekin approach \cite{ADT} and also the ones obtained
by the Hamiltonian procedure in first-order formalism in ref.\cite%
{Blagojevic-Cvetkovic}. The mismatch in the mass respect to the value
obtained by ADT formula ($\mathcal{M}=(1/2)\,\mathcal{M}_{\text{ADT}}$) is
similar to the one found for ACL black holes \cite{ACL}, when the charges
are computed by SAM method \cite{Clement}.


\section{Summary and Outlook}

In this work, we have used the surface terms that come from the first-order
formulation of TMG with negative cosmological constant to construct
background-independent charges from the Noether theorem. It has been shown
that this definition of conserved quantities gives finite results for
locally AdS black holes, solutions of Log Gravity and warped AdS black holes
without the need for any background-substraction procedure.

In ref.\cite{Sezgin-Tanii}, the derivation of TMG from an action that
depends explicitly on the torsion through a constraint imposed by a
Lagrange multiplier has also been employed to obtain the Witten-Nester
charges in the supersymmetric extension of this gravity theory. The
generality of the procedure makes possible a comparison between
Witten-Nester and ADT charges and a further generalization of the latter to
an arbitrary background. This also shows the equivalence with the conserved
quantities discussed in ref.\cite{Clement}. Background-independence is
encoded in the charges obtained by the SAM method, but it is not obvious to
us that one can reconstruct the covariant formulas found here from the
expressions given in a particular frame in ref.\cite{Clement}. In any case,
it would be interesting to explore the relation of the existing methods to
the charges presented in this paper.

In gravity with standard AdS asymptotics, the existence of
background-independent conserved quantities is a consequence of holographic
renormalization, in the sense that this procedure identifies the divergences
in the action and provides a systematic construction of the counterterms
needed to get rid of them.

In TMG, the situation is kind of the opposite as the obtention of
background-independent charges above may be regarded as a step toward a
holographic formulation of the theory.

In that respect, it is clear that the method carried out here should reduce
to the discussion on holographic stress tensor by Kraus-Larsen \cite%
{Kraus-Larsen} and Solodukhin \cite{Solodukhin} for asymptotically locally
AdS spacetimes.

For a modified asymptotic behavior of the metric that includes logarithmic
terms, the finiteness of the Noether charges shown in Subsection \ref%
{LogGravity} suggests that holographic renormalization might be performed in
Log Gravity using a metric frame which considers relaxed Brown-Henneaux (or
Fefferman-Graham) fall-off conditions \cite%
{Grumiller-Johansson1,Grumiller-Johansson2,Maloney-Song-Strominger}.

The case of asymptotically warped AdS$_{3}$ is expected to be both
conceptual and technically more challenging. However, new insight on
boundary conditions for warped AdS black holes (see, e.g., \cite%
{Compere-Detournay bcs}) and recent study of the asymptotic structure of the
field equations for TMG discussed in ref.\cite{Skenderis-Taylor-van Rees}
could make possible a better understanding of the theory in the context of
AdS/CFT correspondence.


\section*{Acknowledgments}

We are indebted to G. Giribet for countless enlightening discussions and
helpful remarks. We also thank B. Cvetkovi\'{c} for introducing us to the
approach described in Section \ref{TMG-first} and S. Detournay for valuable
correspondence. This work was funded by FONDECYT Grants 11070146 and
1090357, Project MECESUP UCV0602 (O.M.) and PUCV Grants 123.797/2007 and
123.702/2009.


\appendix{}

\section{AdS group \label{AdS group}}

The AdS group in three dimensions is isomorphic to the four-dimensional
rotation group $SO(2,2)$ that leaves invariant the quadratic form
\begin{equation}
\eta _{\underline{A}\underline{B}}\,x^{\underline{A}}x^{\underline{B}%
}=-x_{0}^{2}+x_{1}^{2}+x_{2}^{2}-x_{3}^{2}=-\ell ^{2}.
\end{equation}

The group has six generators $J_{\underline{A}\underline{B}}=-J_{\underline{B%
}\underline{A}}$, which can be decomposed as $J_{\underline{A}\underline{B}%
}=\{J_{AB},P_{A}\}$, with $A,B,\ldots =0,1,2$, where $J_{AB}$ are Lorentz
rotations and $P_{A}=J_{A3}$ are AdS translations.

$SO(2,2)$ generators satisfy the algebra%
\begin{equation}
\left[ J_{\underline{A}\underline{B}},J_{\underline{C}\underline{D}}\right]
=\eta _{\underline{B}\underline{C}}J_{\underline{A}\underline{D}}-\eta _{%
\underline{A}\underline{C}}J_{\underline{B}\underline{D}}+\eta _{\underline{A%
}\underline{D}}J_{\underline{B}\underline{C}}-\eta _{\underline{B}\underline{%
D}}J_{\underline{A}\underline{C}}\,,
\end{equation}%
or, in terms of the generators $J_{AB}$ and $P_{A}$,%
\begin{eqnarray}
\left[ J_{AB},J_{CD}\right] &=&\eta _{BC}J_{AD}-\eta _{AC}J_{BD}+\eta
_{AD}J_{BC}-\eta _{BD}J_{AC}\,,  \notag \\
\left[ J_{AB},P_{C}\right] &=&\eta _{BC}P_{A}-\eta _{AC}P_{B}\,, \\
\left[ P_{A},P_{C}\right] &=&J_{AC}\,.  \notag
\end{eqnarray}

The Lie-algebra valued gauge connection $A=A_{\mu }dx^{\mu }$ takes the form%
\begin{eqnarray}
A &=&\frac{1}{2}\,W^{\underline{A}\underline{B}}J_{\underline{A}\underline{B}%
}  \notag \\
&=&\frac{1}{2}\,\omega ^{AB}J_{AB}+\frac{1}{\ell }\,e^{A}P_{A}\,,
\end{eqnarray}%
and the corresponding AdS curvature $F=\frac{1}{2}\,F_{\mu \nu }\,dx^{\mu }%
\mbox{\tiny $\wedge$}dx^{\nu }$ is%
\begin{eqnarray}
F &=&dA+A\mbox{\tiny $\wedge$}A=\frac{1}{2}\,\mathcal{F}^{\underline{A}%
\underline{B}}J_{\underline{A}\underline{B}}  \notag \\
&=&\frac{1}{2}\left( R^{AB}+\frac{1}{\ell ^{2}}\,e^{A}\mbox{\tiny $\wedge$}%
e^{B}\right) J_{AB}+\frac{1}{\ell }\,T^{A}P_{A}\,.
\end{eqnarray}

Bianchi identity for AdS group $\nabla _{\text{AdS}}F=0$ implies the
standard differential and algebraic Bianchi identities%
\begin{equation}
D(\omega )R^{AB}=0\,,\qquad D(\omega )T^{A}=R^{AB}\mbox{\tiny $\wedge$}%
e_{B}\,.
\end{equation}

In a similar fashion as for four-dimensional Lorentz group, there exist two
inequivalent invariant tensors for $SO(2,2)$, that is,%
\begin{eqnarray}
\left\langle J_{\underline{A}\underline{B}}J_{\underline{C}\underline{D}%
}\right\rangle _{1} &=&\epsilon _{\underline{A}\underline{B}\underline{C}%
\underline{D}}\,,  \notag \\
\left\langle J_{\underline{A}\underline{B}}J_{\underline{C}\underline{D}%
}\right\rangle _{2} &=&\eta _{\left[ \underline{A}\underline{B}\right] ,%
\left[ \underline{CD}\right] }\equiv \eta _{\underline{A}\underline{D}}\eta
_{\underline{B}\underline{C}}-\eta _{\underline{B}\underline{D}}\eta _{%
\underline{A}\underline{C}}\,,  \label{InvTensor}
\end{eqnarray}%
whose only non-vanishing components are%
\begin{equation}
\left\langle J_{AB}P_{C}\right\rangle _{1}=\epsilon _{ABC}\,,
\label{IT-epsilon}
\end{equation}%
with the convention $\epsilon _{ABC3}\equiv \epsilon _{ABC}$, and%
\begin{eqnarray}
\left\langle J_{AB}J_{CD}\right\rangle _{2} &=&\eta _{AD}\eta _{BC}-\eta
_{BD}\eta _{AC}\,,  \notag \\
\left\langle P_{A}P_{B}\right\rangle _{2} &=&\eta _{AB}\,.  \label{IT-delta}
\end{eqnarray}%
A different choice of the trace of the AdS generators in a Chern-Simons form
for $SO(2,2)$ group%
\begin{equation}
L_{\text{CS}}(A)=\left\langle AdA+\frac{2}{3}\,A^{3}\right\rangle _{\text{AdS%
}}
\end{equation}%
produces different gravity actions. Indeed, taking (\ref{IT-epsilon}) as the
corresponding invariant tensor, the Chern-Simons density is proportional to
the Lagrangian of standard gravity%
\begin{equation}
\left\langle AdA+\frac{2}{3}\,A^{3}\right\rangle _{1}=\frac{1}{\ell }%
\,\epsilon _{ABC}\,\left( R^{AB}+\frac{1}{3\ell ^{2}}\,e^{A}e^{B}\right)
e^{C}-\frac{1}{2\ell }\,d\left( \epsilon _{ABC}\,\omega ^{AB}e^{C}\right) \,.
\end{equation}%
In the Gauss-normal frame (\ref{Gaussian}), it is easy to see that the
boundary term in eq.(\ref{BoundCS}) is half of the Gibbons-Hawking term,
because%
\begin{equation}
\frac{1}{32\pi G}\,\epsilon _{ABC}\,\omega ^{AB}e^{C}=\frac{1}{16\pi G}\,%
\sqrt{-h}\,K\,.
\end{equation}%
Any asymptotically AdS spacetime metric can be put in a Fefferman-Graham
form \cite{FG}%
\begin{equation}
ds^{2}=\frac{\ell ^{2}}{4\rho ^{2}}\,d\rho ^{2}+\frac{1}{\rho }%
\,g_{ij}\,dx^{i}dx^{j},  \label{FG}
\end{equation}%
where $g_{ij}(x,\rho )$ accepts a regular expansion near the conformal
boundary $\rho =0$%
\begin{equation}
g_{ij}(x,\rho )=g_{(0)ij}+\rho g_{(1)ij}+\cdots \,.
\end{equation}%
The above expansion results in an asymptotic behavior for the extrinsic
curvature
\begin{equation}
K_{j}^{i}=\dfrac{1}{\ell }\,\left( \delta _{j}^{i}-\rho \,k_{j}^{i}\right)
\,,  \label{KFG}
\end{equation}%
with the tensor $k_{j}^{i}$ given by%
\begin{equation}
k_{j}^{i}=g_{(1)ij}+\mathcal{O}(\rho )\,.
\end{equation}%
In three dimensions, Einstein's equation does not fully determine $g_{(1)ij}$
from the initial data $g_{(0)ij}$, but only implies%
\begin{equation}
\nabla _{(0)i}g_{(1)j}^{i}=0  \label{covcon}
\end{equation}
and that the trace is%
\begin{equation}
g_{(1)}=-\frac{\ell ^{2}}{2}\,\mathcal{R}_{(0)}\,.  \label{g1trace}
\end{equation}

It was shown in ref.\cite{Miskovic-Olea3D} that adding and substracting the
Gibbons-Hawking term from the action%
\begin{equation}
I=I_{EH}+\frac{1}{16\pi G}\int\limits_{\partial M}\sqrt{-h}\,K\,,
\end{equation}
one is able to recover the Balasubramanian-Kraus counterterm plus a
topological invariant of the metric $g_{(0)}$. Indeed, with the help of
relations (\ref{FG}-\ref{g1trace}) we have%
\begin{equation}
-\frac{1}{16\pi G}\,\sqrt{-h}\,K=-\frac{1}{8\pi G\ell }\left( \sqrt{-h}+%
\frac{\ell ^{2}}{4}\sqrt{-g_{(0)}}\,\mathcal{R}_{(0)}\right) .
\end{equation}

In turn, the \textit{exotic }copy of CS for AdS group is the sum of the
gravitational Chern-Simons term plus the translational Chern-Simons term $%
e_{A}T^{A}$%
\begin{equation}
\left\langle AdA+\frac{2}{3}\,A^{3}\right\rangle _{2}=\frac{1}{2}\left( L_{%
\text{CS}}(\omega )+\frac{2}{\ell ^{2}}\,e_{A}T^{A}\right) \,.
\end{equation}%
For a given AdS radius, the above relation singles out (\ref{lambda-constant}%
) as the value of the Lagrange multiplier $\lambda _{A}$ that produces a
symmetry-enhancement in the Mielke-Baekler theory \cite{Mielke-Baekler} from
local Lorentz to AdS group in the gravity action (\ref{TMG}).


\section{Gauss-Codazzi relations \label{Gauss-Codazzi}}

We consider a spacetime described by the radial foliation
\begin{equation}
ds^{2}=\mathcal{G}_{\mu \nu }\,dx^{\mu }dx^{\nu
}=N^{2}(r)dr^{2}+h_{ij}(r,x)\,dx^{i}dx^{j}\,,
\end{equation}%
which defines the extrinsic curvature as
\begin{equation}
K_{ij}=\frac{1}{2N}\,h_{ij}^{\prime }\,.
\end{equation}%
In this frame, the non-vanishing components of the Christoffel symbol $%
\Gamma _{\mu \nu }^{\lambda }(\mathcal{G})$ are%
\begin{equation}
\begin{tabular}{ll}
$\Gamma _{rr}^{r}=\dfrac{N^{\prime }}{N}\,,\medskip \qquad $ & $\Gamma
_{ij}^{r}=-\dfrac{1}{N}\,K_{ij}\,,$ \\
$\Gamma _{jr}^{i}=N\,K_{j}^{i}\,,$ & $\Gamma _{jk}^{i}=\Gamma
_{jk}^{i}(h)\,. $%
\end{tabular}%
\end{equation}%
(The metric $h_{ij}$ lowers and raises indices of tensors constructed from $%
h_{ij}$ and $K_{ij}$.) Then, the curvature takes the Gauss-Codazzi form
\begin{eqnarray}
R_{jr}^{ir} &=&-\frac{1}{N}\,\left( K_{j}^{i}\right) ^{\prime }-\left(
K^{2}\right) _{j}^{i}\,, \\
R_{jk}^{ir} &=&-\frac{1}{N}\,\left( \nabla _{j}K_{k}^{i}-\nabla
_{k}K_{j}^{i}\right) \,, \\
R_{kr}^{ij} &=&N\,\left( \nabla ^{i}K_{k}^{j}-\nabla ^{j}K_{k}^{i}\right) \,,
\\
R_{kl}^{ij} &=&\mathcal{R}%
_{kl}^{ij}(h)-K_{k}^{i}K_{l}^{j}+K_{l}^{i}K_{k}^{j}\,,
\end{eqnarray}%
where $\nabla $ is the covariant derivative defined in terms of the
Christoffel symbol of the boundary $\Gamma _{jk}^{i}(h)$ and $\mathcal{R}%
_{kl}^{ij}(h)$ is the Riemann tensor of the boundary.

As a consequence, Ricci tensor and Ricci scalar can be written as
\begin{eqnarray}
R_{j}^{i} &=&\mathcal{R}_{j}^{i}(h)-K_{j}^{i}K-\frac{1}{N}\,\left(
K_{j}^{i}\right) ^{\prime }\,, \\
R_{j}^{r} &=&\frac{1}{N}\,\left( \nabla _{j}K-\nabla _{k}K_{j}^{k}\right) \,,
\\
R_{r}^{r} &=&-\frac{1}{N}\,K^{\prime }-K_{j}^{i}K_{i}^{j}\,,
\end{eqnarray}%
and
\begin{equation}
R=\mathcal{R}(h)-K^{2}-K_{j}^{i}K_{i}^{j}-\frac{2}{N}\,K^{\prime }\,.
\end{equation}


\section{Schouten tensor for stationary metric\ \label{Schouten Appix}}

For a metric of the form
\begin{equation}
ds^{2}=-N^{2}(r)\,dt^{2}+\frac{\ell ^{2}dr^{2}}{4R^{2}(r)N^{2}(r)}%
+R^{2}(r)\left( d\phi +N_{\phi }(r)dt\right) ^{2}\,,
\end{equation}%
the components of the Schouten tensor read%
\begin{eqnarray}
S_{tt} &=&-\frac{1}{2\ell ^{2}}\left( 4N^{4}(R^{\prime
})^{2}+4N^{2}(R^{4})^{\prime }(N_{\phi }^{2})^{\prime }+\frac{1}{2}%
(R^{4})^{\prime }N_{\phi }^{2}(N^{2})^{\prime }-4R^{4}N_{\phi
}^{2}(N^{\prime })^{2}+5R^{6}N_{\phi }^{2}(N_{\phi }^{\prime })^{2}\right.
\notag \\
&&-4R^{4}N_{\phi }^{2}NN^{\prime \prime }+4R^{2}N_{\phi }^{2}(R^{\prime
})^{2}N^{2}+8N^{2}R^{4}N_{\phi }N_{\phi }^{\prime \prime }-\frac{1}{2}%
(N^{4})^{\prime }(R^{2})^{\prime }+4R^{3}N_{\phi }^{2}N^{2}R^{\prime \prime }
\notag \\
&&-\left. 4N^{2}R^{2}(N^{\prime })^{2}-4N^{3}R^{2}N^{\prime \prime
}+3N^{2}R^{4}(N_{\phi }^{\prime })^{2}+4N^{4}RR^{\prime \prime }\right) \,,
\\
S_{t\phi } &=&-\frac{R^{2}}{2\ell ^{2}}(8(R^{2})^{\prime }N^{2}N_{\phi
}^{\prime }+4R^{2}N^{2}N_{\phi }^{\prime \prime }+(R^{2})^{\prime
}(N^{2})^{\prime }N_{\phi }+5R^{4}(N_{\phi }^{\prime })^{2}N_{\phi
}+4(R^{\prime })^{2}N^{2}N_{\phi }  \notag \\
&&+4RN^{2}R^{\prime \prime }N_{\phi }-4N_{\phi }R^{2}(N^{\prime
})^{2}-4N_{\phi }R^{2}NN^{\prime \prime })\,, \\
S_{rr} &=&-\frac{1}{8R^{2}N^{2}}\left( 4R^{2}(N^{\prime
})^{2}+4R^{2}NN^{\prime \prime }-3R^{4}(N_{\phi }^{\prime
})^{2}+4N^{2}RR^{\prime \prime }+4(R^{\prime })^{2}N^{2}+(R^{2})^{\prime
}(N^{2})^{\prime }\right) ,\qquad  \\
S_{\phi \phi } &=&-\frac{R^{2}}{2\ell ^{2}}\left( (R^{2})^{\prime
}(N^{2})^{\prime }+5R^{4}(N_{\phi }^{\prime })^{2}+4(R^{\prime
})^{2}N^{2}+4N^{2}RR^{\prime \prime }-4R^{2}(N^{\prime
})^{2}-4R^{2}NN^{\prime \prime }\right) \,.
\end{eqnarray}

\section{Noether charges and Wald Hamiltonians \label{Wald}}

A Hamiltonian $H(\xi )$ describes the dynamics generated by the vector field
$\xi ^{\mu }$, which is related to Noether current and surface term in eq.(%
\ref{*J}) by
\begin{eqnarray}
\delta H(\xi ) &=&\delta \int\limits_{\partial M}\left. ^{\ast }J\right.
-\int\limits_{\partial M}d\left( \xi \cdot \Theta \right)   \notag \\
&=&\delta \int\limits_{\Sigma _{\infty }}\mathcal{Q}(\xi
)-\int\limits_{\Sigma _{\infty }}\xi \cdot \Theta \,,
\end{eqnarray}%
where we have used the Stokes' theorem for $\left. ^{\ast }J\right. =d%
\mathcal{Q}(\xi )$ to integrate the Noether charge $Q(\xi )=\int_{\Sigma
_{\infty }}\mathcal{Q}(\xi )$. The Hamiltonian exists if there is a $(D-1)$%
-form $\mathcal{B}$ ($D$ is the dimension of spacetime) such that the second
term is a total variation,%
\begin{equation}
\int\limits_{\Sigma _{\infty }}\xi \cdot \Theta (\phi ,\delta \phi )=\delta
\int\limits_{\Sigma _{\infty }}\xi \cdot \mathcal{B}(\phi )\,.
\label{integrability}
\end{equation}%
Only if this is possible, one can write down the Wald Hamiltonian as \cite%
{Iyer-Wald}%
\begin{equation}
H(\xi )=\int\limits_{\Sigma _{\infty }}\left( \mathcal{Q}(\xi )-\xi \cdot
\mathcal{B}\right) \,.  \label{Wald_H}
\end{equation}

It is clear that the integrability criterion (\ref{integrability}) requires
to identify precise boundary conditions, what is not always possible. In
general, the procedure of integrating \ (\ref{integrability}) breaks general
covariance, such that $\mathcal{B}$ is a non-covariant correction to the
charge.

There are, however, other instances where $\mathcal{B}$ can be covariantly
integrated and seen as a contribution coming from a boundary term in the
action, what we show below for standard gravity.

Let us consider Einstein-Hilbert AdS action in three dimensions without
boundary terms,%
\begin{equation}
I_{0}=-\frac{1}{16\pi G}\int\limits_{M}\epsilon _{ABC}\left( R^{AB}+\frac{1}{%
3\ell ^{2}}\,e^{A}e^{B}\right) e^{C}\,,  \label{I0}
\end{equation}%
whose on-shell variation gives a boundary term%
\begin{equation}
\Theta _{0}(\delta e,\delta \omega )=\delta \omega ^{AB}\,\frac{\delta I_{0}%
}{\delta R^{AB}}=-\frac{1}{16\pi G}\,\epsilon _{ABC}\,\delta \omega
^{AB}e^{C}\,.  \label{theta_0}
\end{equation}%
The Noether charge derived from action (\ref{I0}), associated to a Killing
vector is the three-dimensional equivalence of Komar integral, that is,
\begin{equation}
Q_{0}(\xi )=\int\limits_{\Sigma _{\infty }}\xi ^{\mu }\omega _{\mu }^{AB}\,%
\frac{\delta I_{0}}{\delta R^{AB}}=-\frac{1}{16\pi G}\int\limits_{\Sigma
_{\infty }}\epsilon _{ABC}\,\xi ^{\mu }\omega _{\mu }^{AB}e^{C}\,.
\end{equation}%
The Komar formula is obtained in an arbitrary dimension from the dimensional
continuation of (\ref{I0}) and gives $Q_{0}(\partial _{t})=\frac{D-3}{D-2}%
\,M+\lim_{r\rightarrow \infty }\frac{\text{Vol}(S^{D-2})}{8\pi G\ell ^{2}}%
\,r^{D-1}$ for a spherical Schwarzschild-AdS black hole. This means that the
correct mass for the BTZ black hole comes necessarily from the addition of
boundary terms.

We can implement the integrability condition (\ref{integrability}) for the
surface term (\ref{theta_0}) by demanding a boundary condition, whose
fully-covariant version is written as
\begin{equation}
\epsilon _{ABC}\,\delta \omega ^{AB}e^{C}=\epsilon _{ABC}\,\omega
^{AB}\delta e^{C}\,,\qquad \text{ on }\partial M\,.  \label{bcfully}
\end{equation}%
This boundary condition can be explicitly realized in Fefferman-Graham frame
(\ref{FG}-\ref{KFG}) by the conditions $K_{j}^{i}=\frac{1}{\ell }\,\delta
_{j}^{i}+\mathcal{O}(\rho )$ and $\delta K_{j}^{i}=\mathcal{O}(\rho )$,
which are compatible with asymptotically AdS spacetimes \cite%
{Miskovic-Olea3D}.

Using (\ref{bcfully}), we can integrate $\mathcal{B}$ in eq.(\ref%
{integrability}) in the following way,
\begin{eqnarray}
\xi \cdot \Theta _{0}(\delta e,\delta \omega ) &=&-\frac{1}{16\pi G}\,\frac{1%
}{2}\,\xi \cdot \left( \epsilon _{ABC}\,\delta \omega ^{AB}e^{C}+\epsilon
_{ABC}\,\omega ^{AB}\delta e^{C}\right) +\mathcal{O}(1)  \notag \\
&=&-\,\delta \left[ \xi \cdot \left( \frac{1}{32\pi G}\,\epsilon
_{ABC}\,\omega ^{AB}e^{C}\right) \right] \,+\mathcal{O}(1),  \label{varB}
\end{eqnarray}%
and therefore, $\mathcal{B}$ is equal to (up to a finite contribution) minus the
boundary term (\ref{BoundCS}) that regularizes the Einstein-Hilbert part of
the action,
\begin{equation}
\mathcal{B}=-B+\mathcal{O}(1)=-\frac{1}{32\pi G}\,\epsilon _{ABC}\,\omega
^{AB}e^{C}+\mathcal{O}(1).  \label{BequalB}
\end{equation}

A similar procedure can be carried out in four-dimensional AdS gravity in order
to integrate out the boundary term necessary for the regularization of the conserved quantities,
 which also provides a fully-covariant correction to the Komar charge \cite{OleaEven}.

The additional term in (\ref{varB}) is responsible for a finite difference
between the Noether charge $Q_{1}(\xi )$ in eq(\ref{Q1_form}) and the
corresponding Hamiltonian, because%
\begin{equation}
\delta H_{1}(\xi )=\delta Q_{1}(\xi )-\int\limits_{\Sigma _{\infty }}\xi
\cdot \Theta _{1}  \label{deltaH1}
\end{equation}%
where $\Theta _{1}$ is the first part of the surface term (\ref{theta}) or,
equivalently, $\Theta _{1}=\Theta _{0}+\delta B$.

To this end, let us calculate the finite contribution of $\Theta _{1}(\delta
e,\delta \omega )$
\begin{eqnarray}
\Theta _{1}(\delta e,\delta \omega ) &=&-\frac{1}{32\pi G}\,\epsilon
_{ABC}\,\left( \delta \omega ^{AB}e^{C}-\omega ^{AB}\delta e^{C}\right)
\notag \\
&=&-d^{2}x\,\frac{\sqrt{-h}}{16\pi G}\,\left[ \left( K_{j}^{i}-\frac{1}{2}%
\,K\delta _{j}^{i}\right) \left( h^{-1}\delta h\right) _{i}^{j}+\delta K%
\right] \,,
\end{eqnarray}%
and, using expansion of the fields in FG frame (\ref{FG}-\ref{g1trace}), we
find%
\begin{eqnarray}
\Theta _{1} &=&d^{2}x\,\frac{\sqrt{-g}}{16\pi G\ell }\,\left(
k_{j}^{i}-k\delta _{j}^{i}\right) \left( g^{-1}\delta g\right)
_{i}^{j}+d^{2}x\,\frac{1}{16\pi G\ell }\,\delta \left( \sqrt{-g}k\right)
\notag \\
&=&\frac{1}{16\pi G\ell }d^{2}x\,\left[ H_{j}^{i}\left( g_{(0)}^{-1}\delta
g_{(0)}\right) _{i}^{j}-\frac{\ell ^{2}}{2}\,\delta \left( \sqrt{-g_{(0)}}%
\,\mathcal{R}_{(0)}\right) \right] \,,
\end{eqnarray}%
where we have defined the tensor $H_{j}^{i}=g_{(1)j}^{i}-g_{(1)}\delta
_{j}^{i}$ which has conformal weight 2 and, by virtue of eq.(\ref{covcon}),
is covariantly conserved. The second term is a two-dimensional topological
invariant. In this respect, $H_{j}^{i}$ gives rise to an ambiguity in the
definition of the Wald Hamiltonian and thus, to a difference between the
Noether charge $Q_{1}(\xi )$ and the Hamiltonian $H_{1}(\xi )$.

More generally, it has been proved in ref.\cite{Skenderis-Papadimitriou}
that for asymptotically AdS spacetimes in $D=2n+1$ dimensions, the ambiguity
in the definition of a Wald Hamiltonian can be written as%
\begin{equation}
H^{\prime }(\xi )=H(\xi )+\int\limits_{\Sigma _{\infty }}d^{2n-1}y\sqrt{%
\sigma }u_{i}\,H_{j}^{i}\,\xi ^{j}\,,
\end{equation}%
in the notation of the ADM foliation (\ref{ADMh}). The tensor $H_{j}^{i}$
has conformal weight $2n$\ and its trace is proportional to the Weyl anomaly
of the holographic stress tensor. The best known example is the
four-dimensional one, where $H_{j}^{i}$ is quadratic in the curvature
\cite{Skenderis-Papadimitriou,Ashtekar-Das,Hollands-Ishibashi-Marolf}.

In addition to Einstein-Hilbert action, TMG contains the terms%
\begin{equation}
I_{2}=\frac{1}{32\pi G\mu }\int\limits_{M}\left( L_{\text{CS}}(\omega
)+2\,\lambda _{A}T^{A}\right) \,.  \label{I2}
\end{equation}%
For the present discussion on Wald Hamiltonians, we will restrict ourselves
to the asymptotically AdS sector, where $\lambda _{A}=e_{A}/\ell ^{2}$.
Precise boundary conditions suitable for Log Gravity and Warped AdS
asymptotics are currently under investigation.

In this case, the contribution to the surface term of the action (\ref{I2})
is%
\begin{eqnarray}
\Theta _{2} &=&\frac{1}{32\pi G\mu }\,\left( \delta \omega ^{AB}\omega _{BA}+%
\frac{2}{\ell ^{2}}\,\delta e^{A}e_{A}\right)   \notag \\
&=&d^{2}x\,\frac{1}{16\pi G\mu }\,\left( -\delta
K_{i}^{k}K_{kj}-K_{i}^{k}K_{j}^{l}\,e_{al}\delta e_{k}^{a}+\frac{1}{\ell ^{2}%
}\,\delta e_{i}^{a}e_{aj}\right) \epsilon ^{ij}\,,
\end{eqnarray}%
where $e_{i}^{a}$ is the boundary zweibein of the induced metric $%
h_{ij}=\eta _{ab}\,e_{i}^{a}e_{j}^{b}$. The above variation is not
expressible in terms of variations of the metric tensor, but adopts the form
\begin{equation}
\Theta _{2}=d^{2}x\,\tilde{H}_{a}^{i}\,\delta e_{(0)i}^{a}\,,
\end{equation}%
where%
\begin{equation}
\tilde{H}_{a}^{i}=\frac{1}{8\pi G\ell ^{2}\mu }\,g_{(1)j}^{k}\,e_{(0)ak}%
\epsilon ^{ij}
\end{equation}%
is a covariantly conserved tensor with vanishing trace. Indeed, the tensor $%
\tilde{H}^{ij}=2\tilde{H}_{a}^{i}e_{(0)}^{aj}$ has physical relevance,
because it is the same as the holographic stress tensor associated to the
action (\ref{I2}), whose antisymmetric part carries the gravitational
anomaly \cite{Solodukhin}
\begin{equation}
\tilde{H}^{ij}\epsilon _{ij}=\frac{1}{8\pi G\mu }\,\mathcal{R}_{(0)}\,.
\end{equation}%
In an analogous way as for the Einstein-Hilbert action, where Weyl anomaly
is nonvanishing, one could expect an ambiguity in the Wald Hamiltonian
definition due to the existence of gravitational anomaly.


\end{document}